# Study on the Concept and Development of a Mobile Incubator

Fehmi Can Ay[1*], Nesim Bilici[1*], Rahmetullah Varol[1], Atasangu Yilmaz[2], Ufuk Gorkem Kirabali[1], Abdurrahim Yilmaz[1], Huseyin Uvet[1**]

[1]Mechatronics Engineering Department, Yildiz Technical University, Istanbul 34349, Turkey

[2]Uskudar American Academy, Istanbul, Turkey

*Authors have equal contributions.

**Corresponding author: Huseyin Uvet (huvet@yildiz.edu.tr)

## Abstract

Creating the best possible conditions is essential for proper cell growth. Incubators, a type of biotechnological instrument, are used to simulate this condition and maintain the cells within them. The processes involved in creating a mobile incubator, which are essential for monitoring a cell culture's physiological parameters, are outlined in this article. The goal is to keep image-taking during cell development from compromising data accuracy. The cell culture is prone to contamination once it has been removed from the incubation environment for further monitoring. The proposed approach allows for on-the-go monitoring of the cell culture. Moreover, it enables constant monitoring.

## 1. Introduction

The incubator has served several functions in the lab throughout the years, from facilitating the hatching of chicken eggs to helping researchers learn about and develop vaccines against dangerous pathogens [1]. The incubator has also aided in cellular and molecular biology research, leading to medical advances [2].

An incubator consists of a temperature-controlled chamber. Within the incubator, certain incubators regulate humidity, gas composition, and ventilation. While much has changed technologically since the first incubators were employed in ancient Egypt and China [3], the incubator's basic goal has remained the same: to provide a stable, regulated environment suitable to research, study, and cultivation.

The introduction of the CO2 incubator marked a significant advancement in incubator design [4]. The demand for CO2 incubators increased as medical professionals realized they could use them to study bacteria and viruses isolated from patient fluids. A sample was collected and cultured in an incubator using a clean Petri dish. The carbon dioxide and nitrogen contents in the incubator were set at the levels required to drive cell growth, and the air temperature was maintained at 37 degrees Celsius, the same as a human body.

Incubators were first utilized in genetic engineering about this time. With the help of incubators, scientists could produce biologically important proteins like insulin. On a molecular level, genetic manipulation is now possible.

Incubators were also being employed in genetic engineering at this time. Incubators allowed scientists to produce insulin and other biologically significant proteins. Genetic modification can now be used at the cellular level to make fruits and vegetables healthier and more resistant to diseases and bugs.

The incubator serves multiple functions in a research laboratory. While its primary role is temperature control, modern incubators also have a number of other uses. Humidity settings are a common addition to incubators [5]. Many incubators designed for laboratory use feature a backup power supply to guarantee that experiments are not disrupted in the event of a power outage. Incubators come in a wide range of sizes, from tabletop units to warm rooms that can hold a huge number of samples.

Even though it is crucial for researchers to incubate living cells, observing how the cells adapt to their new environment is even more significant. Cell images are captured using various techniques for this purpose [6] [7]. During image capture, the cell is taken out of its optimal environment and put in a place where it could be contaminated. The term "optimal environment" refers to the artificially humidified atmosphere that is manufactured to promote cell growth [8]. This causes a discrepancy in the precision of the information that has been received. By simulating the ideal conditions in a compact, transportable, and imaging-enabled device, the proposed mobile incubator seeks to improve the reliability of collected data.

## 2. Design

The controller's sensors send information to the mobile incubator's core working logic, which in turn triggers the device's actuators. By doing so, it simulates the conditions inside a cell to produce what is known as the "optimal environment." The feedback loop regulates and stabilizes the cellular environment in real time. When a live cell is inserted in the device and begins to develop, an image can be captured from the device using the appropriate image acquisition method in order to study the cell without removing it from the optimal environment. This image can be captured and studied later on a computer, or it can be examined in real time. Figure 1 depicts the system's detailed working scheme.

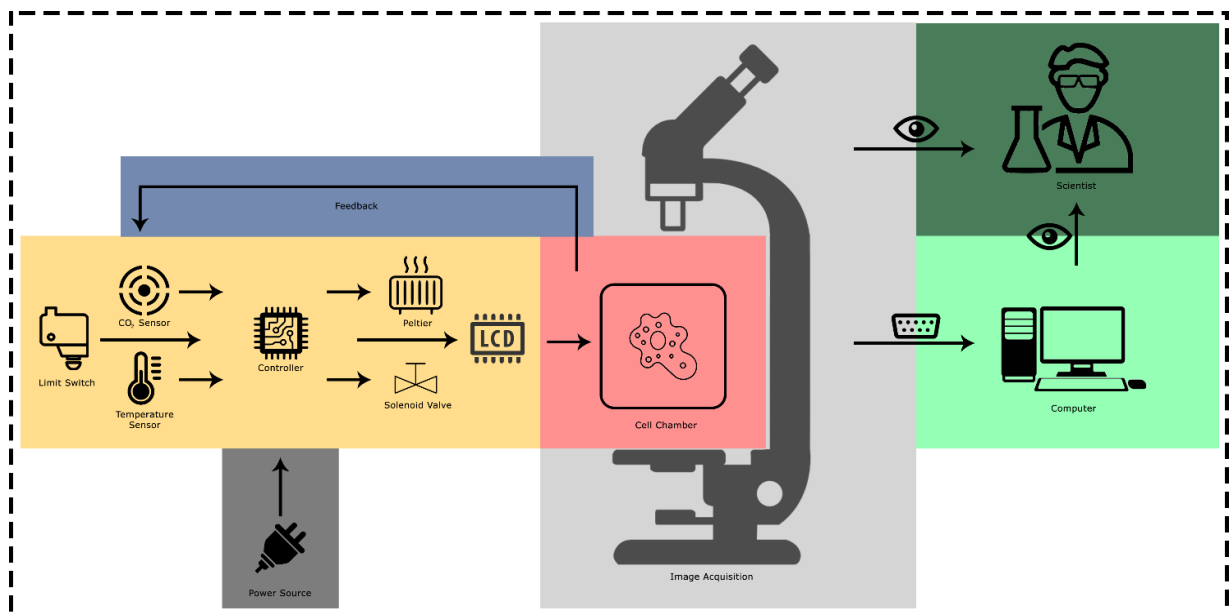

*Figure 1: System Working Scheme*

During the design phase of the device, some limitations were determined by considering the knowledge and experience obtained in the context of research. These constraints are that the flask in which the cell will be transported fits into the cell section of the device; the weight of the device is suitable for portability; the material from which the device will be produced does not react positively or negatively on the cells; the device can withstand the vacuum effect that will occur inside when the desired conditions are met; the size of the device is small enough to be able to fit into the device where the image is to be taken; and the cell chamber is completely sealed separately from the electrical part. The design was carried out by adhering to these 6 constraints as shown in Figure 2. There are 2 different parts in the design, as seen in Figure 3. These parts are the control panel part and the cell chamber part, respectively. There is a spacer that separates these two parts from each other. The sensors, which perform the device's most important function, are mounted on the wall in between. So, with the help of sensors, the controller can get the data from the chamber and figure out what to do with it.

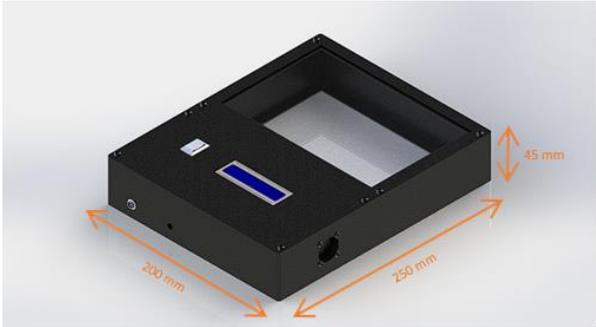

Figure 2: 3D Design and Size

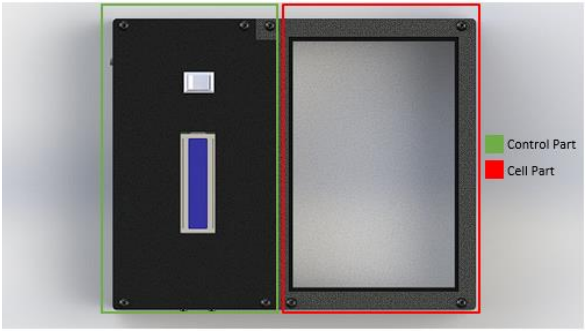

Figure 3: Control and Cell Chamber Areas

## 3. Analysis

The mobile incubator is modeled as an enclosed space to study its thermal properties. The on-off system managed the smart mobile incubation station's streamlined closed-box heating system [9]. Sensor data on the system's temperature will trigger the activation of the heater whenever that temperature drops below a predetermined threshold. When needing a quick assessment of the heating and cooling system's efficiency, a first-order model is applied. The formula below was used to determine the system's transfer function.

$$hA[(x - x_s) - (y - y_s)] = mc_p \frac{d(y - y_s)}{dt}, \quad [10]$$

$$Note \; \frac{d(y - y_s)}{dt} = \frac{dy}{dt}$$

$$[X - Y] = \frac{mc_p}{hA} \cdot \frac{dy}{dt}$$

*By applying Laplace Transform;*

$$[X(s) - Y(s)] = \frac{mc_p}{hA} sY(s)$$

$$\frac{Y(s)}{X(s)} = \frac{1}{\tau s + 1}, \quad \tau = \frac{\rho V c_p}{hA}$$

$$\rho = density \left(\frac{kg}{m^3}\right)$$

$$V = volume \ (m^3)$$

$$m = mass \ \ (kg)$$

$$c_p = specific\ heat \ \left(\frac{J}{kg \cdot K}\right)$$

$$h = Convective\ heat\ transfer\ coefficent \left(\frac{W}{m^2 \cdot K}\right)$$

$$A = Cross\ sectional\ area\ perpandicular\ to\ heat\ flow\ (m^2)$$

Variables for air between 293K and 303K ;

$\rho = 1.184, \quad c_p = 1007, \quad h = 10$

$V = 0.00065, \quad , \quad A = 0.0036$

$$\tau = \frac{\rho V c_p}{hA}$$

$$\tau = 21.53\ sec$$

So transfer function becomes;

$$TF = \frac{1}{21.53s + 1}$$

The modeling of the system was created on MATLAB-Simulink as seen in Figure 4.

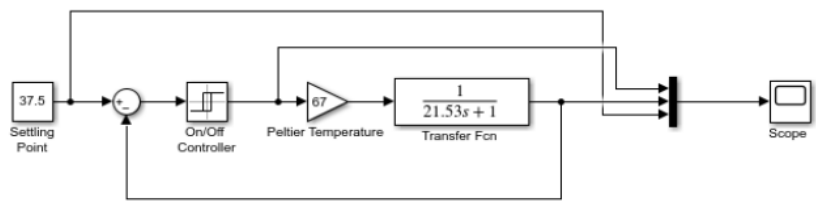

*Figure 4: MATLAB- Simulink Model of System*

The heating activity of the system during operation was obtained as seen in Figure 5.

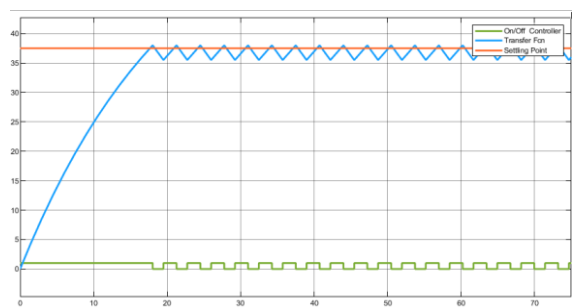

*Figure 5: System Working Graph*

# 4. Simulation

Research has shown that a concentration of 5% $CO_2$ in the surrounding atmosphere is necessary for optimal cell development [11]. The necessary amount of carbon dioxide was supplied by the carbon dioxide tube, and the gas was released into the atmosphere after passing through a syringe filter. The solenoid valve regulates the amount of fluid that exits the tube. An ANSYS finite element model was used to simulate the chamber to be utilized in the design. The simulation found that the imbalance in the chamber results in certain vortices, represented by the red circles in Figure 6.

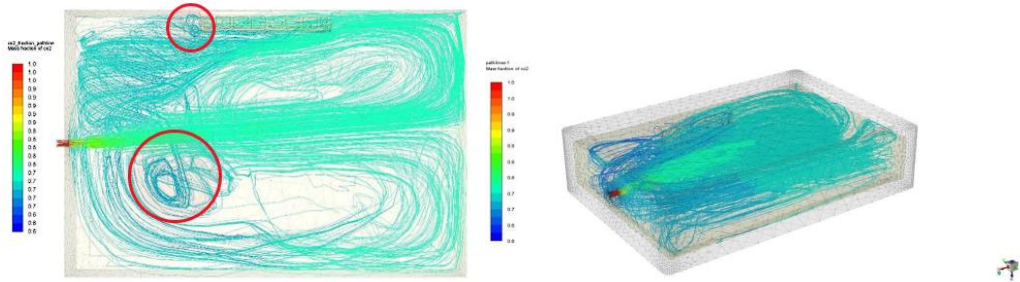

*Figure 6: Flow Simulation in ANSYS*

During the simulation, it has been determined that the carbon dioxide given to the chamber reaches a maximum speed of ≈10 m/s while passing through a 3 mm diameter hole under constant pressure.

# 5. Implementation

The smart mobile incubator station has an aluminum 7000 series exoskeleton. This choice was made with several factors in mind, including longevity, portability, thermal conductivity, and impermeability. Figure 7 displays the physical characteristics of aluminum from the 7000 series [12].

| Physical Properties | Metric |
|---|---:|
| Density | 2.72 - 2.89 g/cc |
| **Mechanical Properties** | **Metric** |
| Hardness, Brinell | 20.0 - 210 |
| Hardness, Knoop | 80.0 - 232 |
| Hardness, Rockwell A | 39.1 - 59.0 |
| Hardness, Rockwell B | 50.0 - 96.0 |
| Hardness, Vickers | 68.0 - 217 |
| Tensile Strength, Ultimate | 70.0 - 750 MPa |
| Tensile Strength, Yield | 69.0 - 730 MPa |
| Elongation at Break | 1.00 - 25.0 % |
| Creep Strength | 434 - 538 MPa |
| Rupture Strength | 469 - 552 MPa |
| Modulus of Elasticity | 67.0 - 73.0 GPa |
| Compressive Yield Strength | 372 - 669 MPa |
| Compressive Modulus | 70.0 - 72.4 GPa |
| Bearing Yield Strength | 538 - 910 MPa |
| Poissons Ratio | 0.330 |
| Fatigue Strength | 140 - 425 MPa |
| Fracture Toughness | 16.5 - 150 MPa-m½ |
| Machinability | 70.0 - 90.0 % |
| Shear Modulus | 25.0 - 27.6 GPa |
| Shear Strength | 50.0 - 400 MPa |
| **Electrical Properties** | **Metric** |
| Electrical Resistivity | 0.00000290 - 0.00000590 ohm-cm |
| **Thermal Properties** | **Metric** |
| CTE, linear | 21.4 - 25.5 µm/m-°C |
| Specific Heat Capacity | 0.856 - 0.960 J/g-°C |
| Thermal Conductivity | 115 - 222 W/m-K |
| Melting Point | 476 - 657 °C |
| Solidus | 476 - 641 °C |
| Liquidus | 627 - 657 °C |

*Figure 7: Properties of Aluminum 7000 Series*

The skeleton system consists of multiple aluminum parts produced by laser cutting and machining methods. Assemble of these parts was made with a polyurethane-based adhesive. Technical properties of polyurethane-based adhesive are shown in Figure 8 [13] .

| **Technical Properties** | |
|---|---|
| Density | 1.13 ± 0.03 gr/ml |
| Color | Aluminum |
| Basis | Polyurethane Prepolymer |
| Tack-Free Time | 5 – 10 min. (at 23°C and %50 R.H.) |
| Consistency | Thixotropic |
| Shrinkage | None |
| Temperature Resistance | -20°C to +70°C |
| Application Temperature | +5°C to +35°C |
| Maximum Shear Strength | (beech-beech) |
| After 15 min | > 50 kgf/cm2 |
| After 24 hours | > 80 kgf/cm2 |

*Figure 8: Technical Properties of polyurethane-based adhesive*

This glue helped to ensure the sealing of the cell chamber as well as the construction of the smart mobile incubator station.

# 6. Experimental Results

The results of the experiments demonstrated that the sensors were capable of accurately measuring the ambient temperature and carbon dioxide concentration. Data was compared to reference values determined at room temperature, and then the data was calibrated.

In low-light tests, it was found that adequate light was transmitted through the borosilicate glass and into the CMOS camera. For a holographic imaging experiment, we also looked at developing laser illumination using a 10mW green source (532nm), and the resulting device could let laser light pass through and produce usable fringe patterns.

It has been determined that none of the software or hardware components of the system have any abnormalities after conducting tests on each one separately. Figure 9 and Figure 10 show images of cell counting and wound opening obtained by placing the mobile incubator system beneath an inverted microscope for image acquisition.

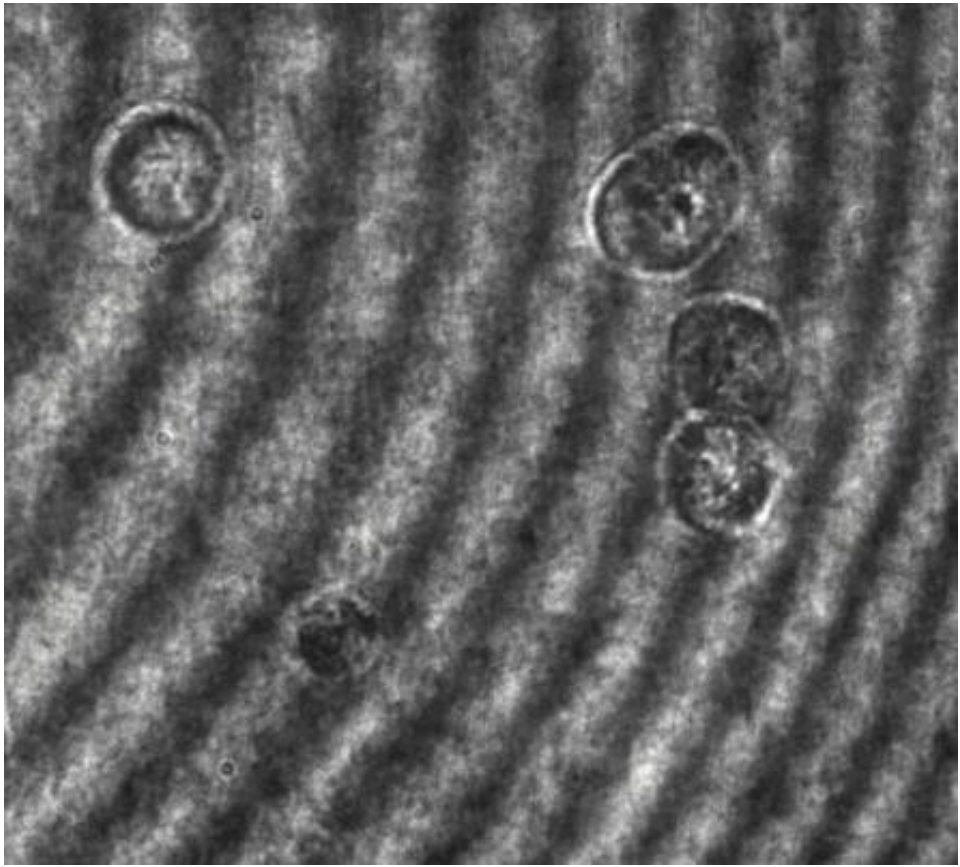

*Figure 9: Cell Counting Image Acquisition*

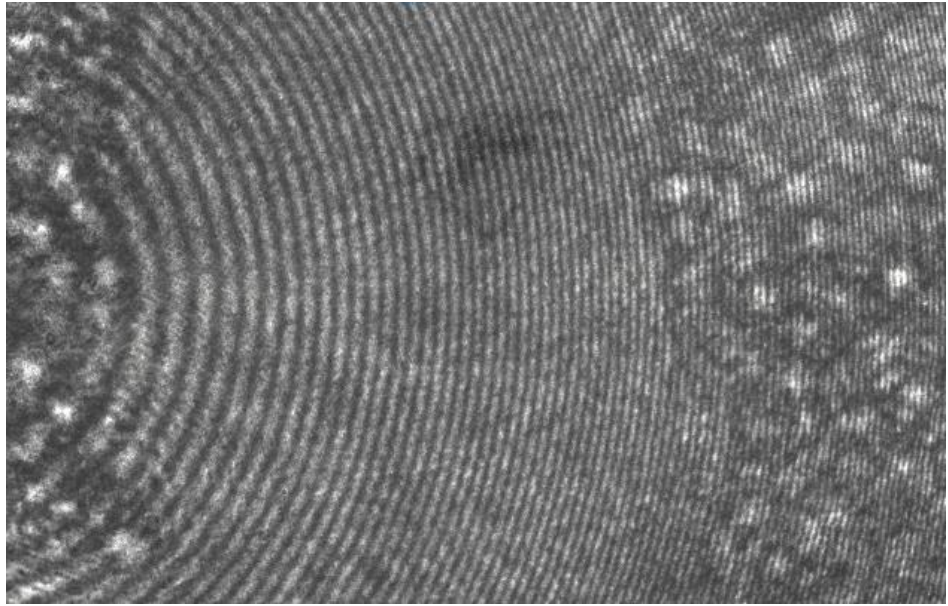

*Figure 10: Wound Opening Image Acquisition*

# 7. Conclusion

A smart mobile incubation station that permits both picture acquisition and mobility has been examined, along with its underlying working concept and experimental results, in this research. With the aid of a heating pad and carbon dioxide tube, the incubator system met its minimal requirements of 37.5% temperature and 5% carbon dioxide rate. Carbon dioxide gas was introduced into the system via a solenoid air valve, which allowed for precise regulation of the gas's flow. The on/off control algorithm described in section 3 operates the valve and heating pad. Borosilicate glass is used for the cell's floor and ceiling because of its low softening temperature (800 C), low coefficient of thermal expansion ($\sim(3-6)x10^{-6}K^{-1}$), and capacity to permit the image acquisition [14]. Aluminum was chosen for the smart incubator system's enclosure due to its durability, low weight, and high thermal conductivity.

The system was evaluated in a laboratory setting using human epithelial and ovarian cells as the subjects of the tests. The findings that were discovered during the trials are broken down in significant detail in section 4. The data gathered from the smart incubator's experiments can then be used to inform next steps for the user. As a consequence of the trials that were carried out, the smart incubator system gives users the possibility to carry out image acquisition and mobility without taking the cells away from their optimal surroundings.